\documentclass[11pt]{article}
\usepackage{amssymb,latexsym,amsmath, amsthm}
\usepackage{amsfonts, graphics}
\def\R{\mathbb R}

\def\supp{\mathrm{supp}\,}
\def\div{\mathrm{div}\,}
\def\dt{\partial_t}
\def\dx{\partial_x}

\def\dv{\partial_v}
\newcommand{\prf}{{\bf Proof.\ }}
\newcommand{\prfe}{\hspace*{\fill} $\Box$

\smallskip \noindent}

\newtheorem{theorem}{Theorem}[section]

\newtheorem{lemma}[theorem]{Lemma}
\theoremstyle{remark}
\newtheorem*{remark}{Remark}

\begin{document}

\title{Gravitational collapse and the \\ Vlasov-Poisson system}
\author{Gerhard Rein, Lukas Taegert\\
        Fakult\"at f\"ur Mathematik, Physik und Informatik\\
        Universit\"at Bayreuth\\
        D-95440 Bayreuth, Germany\\
        email: gerhard.rein@uni-bayreuth.de, lukas.taegert@uni-bayreuth.de}

\maketitle

\begin{abstract} 
A self-gravitating homogeneous ball of a fluid with pressure
zero where the fluid particles are initially at rest collapses
to a point in finite time. We prove that this gravitational collapse 
can be approximated arbitrarily closely by suitable solutions
of the Vlasov-Poisson system which are known to exist globally in time.
\end{abstract}

\maketitle

\section{Introduction}
\setcounter{equation}{0}

Perhaps the simplest example of a matter distribution which collapses under the
influence of its own, self-consistent gravitational field is 
a homogeneous ball of an ideal, compressible fluid with the equation of
state that the pressure is identically zero---this is usually referred to as 
dust---and with the particles initially being at rest. In suitable
units, the radius $r(t)$ of this ball is determined by the initial value problem
\[
\ddot r = -\frac{1}{r^2},\ r(0)=1,\ \dot r(0)=0.
\]
The mass density is given by
\[
\rho(t,x) = \frac{3}{4\pi} \frac{1}{r^3(t)} \mathbf{1}_{B_{r(t)}(0)}(x),\ 
t\geq 0,\ x\in \R^3.
\]
Here $\mathbf{1}_S$ denotes the indicator function of the set $S$ and
$B_r(z)$ is the ball of radius $r>0$ centered at $z\in \R^3$. 
There exists a time $T>0$ such that
$r(t)>0$ exists on $[0,T[$ with 
$\lim_{t\to T} r(t)=0$, i.e., the dust ball collapses to a point in finite time. 
It is easy to check that the above mass density solves the pressure-less 
Euler-Poisson system
\begin{equation} \label{eulercont}
\dt \rho + \div (\rho u) = 0,
\end{equation}
\begin{equation} \label{eulernewton}
\dt u + (u\cdot \dx) u = -\dx U,
\end{equation}
\begin{equation}
\Delta U = 4\pi \rho,\ \lim_{|x| \to \infty} U(t,x) = 0, \label{poisson}
\end{equation}
with velocity field 
\[
u(t,x)=\frac{\dot r(t)}{r(t)} x;
\]
$U=U(t,x)$ denotes the induced gravitational potential.
The fact that such a matter distribution collapses is not surprising
since there is no mechanism which opposes gravity, the pressure being
put to zero by the choice of the equation of state. This situation changes 
significantly if dust is replaced by a collisionless gas as matter model
in which case smooth, compactly supported initial data launch solutions
which exist globally in time and do not undergo a gravitational collapse
in the above sense, cf.~\cite{LP,Pf}. 
As in dust, the particles in a collisionless gas interact only by
gravity, but the particle ensemble is now given in terms
of a density function $f=f(t,x,v)\geq 0$ on phase space where
$t\in \R,\ x,v\in \R^3$ stand for time, position, and velocity,
and $f$ obeys the
Vlasov-Poisson system which consists of the Vlasov equation
\begin{equation}
\partial_{t}f+v\cdot \dx f - \dx U\cdot \dv f =0,
\label{vlasov}
\end{equation}
coupled to the Poisson equation \eqref{poisson} via the definition
\begin{equation}
\rho(t,x) = \int f(t,x,v)\,dv \label{rhodef} 
\end{equation}
of the spatial mass density in terms of the phase space density $f$;
unless indicated otherwise, integrals always extend over $\R^3$.
In astrophysics, the Vlasov-Poisson system 
\eqref{poisson}, \eqref{vlasov}, \eqref{rhodef} is used as a model for
galaxies or globular clusters, cf.~\cite{BT}.
The relation between its solutions and those of the pressure-less 
Euler-Poisson system
\eqref{eulercont}, \eqref{eulernewton}, \eqref{poisson}
was investigated in \cite{DS}. If $(\rho,u)$ is a solution of
the former system, then formally $f(t,x,v)=\rho(t,x)\, \delta(v-u(t,x))$
satisfies the latter system where $\delta$
denotes the Dirac distribution. However, when we speak
of the Vlasov-Poisson system in the present paper 
we only consider genuine functions
on phase space (which will actually be smooth) as solutions,
and we ask the question whether the collapsing solution of the
pressure-less Euler-Poisson system
presented at the beginning of this introduction can be approximated
by suitable smooth solutions of the Vlasov-Poisson system.
This is indeed possible, cf.~Theorem~\ref{thm:approx} below,
and yields the existence of nearly
collapsing solutions of the Vlasov-Poisson system. 

To make the latter more
precise we recall the definitions of the kinetic
and the potential energy associated with a solution:
\begin{eqnarray*}
E_\mathrm{kin}(t)
&=& \frac{1}{2} \iint |v|^2 f(t,x,v)\, dv\, dx,\\
E_\mathrm{pot}(t)
&=& - \frac{1}{8\pi} \int |\dx U(t,x)|^2 dx
= \frac{1}{2}\iint \frac{\rho(t,x)\, \rho(t,y)}{|x-y|}dy\,dx.
\end{eqnarray*}
The total energy $E_\mathrm{kin}(t) + E_\mathrm{pot}(t)$
is conserved.
We also recall that a solution of the Vlasov-Poisson system is
{\em spherically symmetric} if
$f(t,x,v)=f(t,Ax,Av)$ for any rotation $A\in \mathrm{SO}(3)$.
For a spherically symmetric solution and by abuse of notation,
$\rho(t,x)=\rho(t,r)$ and 
\[
\dx U(t,x) = \frac{m(t,r)}{r^2}\frac{x}{r},\ |x|=r,
\]
where
\[
m(t,r)=4\pi\int_0^r s^2\rho(t,s)\, ds
\]
is the mass contained in the ball of radius $r>0$ centered at the origin.
It is well known that spherically symmetric initial data launch spherically
symmetric solutions of the Vlasov-Poisson system, cf.~\cite{Rein07}.
\begin{theorem}\label{thm:Main}
For any constants $C_1, C_2>0$ there exists a smooth,
spherically symmetric solution 
$f$ of the Vlasov-Poisson system such that
initially
\[
||\rho(0)||_\infty ,\ E_\mathrm{kin}(0),\  - E_\mathrm{pot}(0),\
\sup_{r > 0} \frac{m(0,r)}{r}
\leq C_1,
\]
but for some time $t>0$,
\[
||\rho(t)||_\infty,\  E_\mathrm{kin}(t),\
- E_\mathrm{pot}(t),\ \sup_{r > 0} \frac{m(t,r)}{r} > C_2.
\]
\end{theorem}
By choosing $C_1$ small and $C_2$ large
we see that the matter distribution is initially dilute
and the particles are nearly at rest, but at some later time $t$
the matter is very concentrated with large total kinetic and potential energy.
In this sense, Vlasov-Poisson solutions can be very close to a gravitational
collapse even though---as opposed to the case of the pressure-less
Euler-Poisson system---the quantities in the above theorem
remain bounded on any bounded time interval. 
Besides the wish to understand better the relation
of the two systems under consideration there is a more specific
motivation for the current investigation which also explains
why the term $m/r$ is considered in the theorem above.
This motivation originates in general relativity.

In 1939, J.~R.~Oppenheimer and H.~Snyder \cite{OS} showed 
how a black hole can develop from regular data. 
Much as in our introductory example, they considered
a spherically symmetric, asymptotically flat spacetime with 
a dilute homogeneous ball of dust as matter model and showed that 
a trapped surface and hence a black hole form in the evolution.
If $r$ denotes the area radius, then the condition $2 m/r >1$
indicates that the sphere of radius $r$ is trapped 
and the spacetime contains a black hole. Here $m$
is the appropriate general relativistic analogue of the mass
function introduced above, the so-called quasilocal
ADM mass, and $m(t,r=\infty)$ is a conserved quantity,
the ADM mass.
The Oppenheimer-Snyder example suffers from the fact that
there is no pressure in the matter model, and
it remains unclear if a similar calculation 
is possible with a matter model which is not pressure-less.
The analysis in the present note is intended as a blue-print
for the analogous analysis in the general relativistic setting
which will lead to Oppenheimer-Snyder type solutions which collapse
to a black hole,
but with the physically more realistic Vlasov equation as a matter model,
cf.~\cite{diss}.

Both the Oppenheimer-Snyder solution 
and its Newtonian analogue can be obtained by taking a spatially homogeneous
solution with a big crunch singularity in the future, cutting a suitable, 
spatially finite piece from it and extending it by vacuum. If the cutting is 
done along the trajectory of a dust particle, a consistent, asymptotically 
flat solution of the desired form is obtained. In the present analysis we 
follow the same recipe.
We first introduce a class of spatially homogeneous,
cosmological solutions of the Vlasov-Poisson system with a singularity 
in the future;
such solutions, which do of course not
satisfy the boundary condition in \eqref{poisson},
and their perturbations were considered in \cite{RR}.
From such a spatially homogeneous solution we cut a ball centered at the origin
at time $t=0$ and extend it smoothly by vacuum. This provides the initial
data for the Vlasov-Poisson solution. The latter will for some time have
a spatially homogeneous region at the center, the size of which we can 
control. By choosing the original homogeneous solution sufficiently close 
to a dust solution in a suitable sense, the time for which the homogeneous 
core persists can be pushed as close to the collapse time of the homogeneous 
solution as we wish. This will prove Theorem~\ref{thm:approx} from which 
Theorem~\ref{thm:Main} will follow.

\section{Solutions with a homogeneous core}
\setcounter{equation}{0}

We first recall the construction of spatially
homogeneous solutions to the Vlasov-Poisson system.
To do so we fix a continuously differentiable function
$H\colon \R\to [0,\infty[$ 
with support $\supp H\subset [0,1]$ and
\[
\int H(|v|^2)\, dv = \frac{3}{4\pi}.
\]
For $\epsilon\in ]0,1]$ let
\begin{equation}\label{eq:DefinitionHepsilon}
H_\epsilon:=\frac{1}{\epsilon^3}H\left(\frac{\cdot}{\epsilon^2}\right)
\end{equation}
so that 
$\supp H_\epsilon\subset [0,\epsilon^2]$ and 
\[
\int H_\epsilon(|v|^2)\, dv = \frac{3}{4\pi}.
\]
Let $a \colon  [0,T[ \to ]0,\infty[$ be the maximal solution of
\begin{equation} \label{eq:Definitiona}
\ddot a=-\frac{1}{a^2},\
a(0)=\mathring a,\quad \dot a(0)=0,
\end{equation}
where $\mathring a >0$ is prescribed.
A straight forward computation shows that
\begin{equation} \label{hdef}
h_\epsilon \colon [0,T[\times\R^3\times\R^3\to[0,\infty[, \quad 
h_\epsilon(t,x,v):= H_\epsilon(|a(t)v-\dot a(t) x|^2)
\end{equation}
is a spherically symmetric solution of the Vlasov-Poisson system---where
the boundary condition at spatial infinity is dropped---with
\[
\rho_h(t,r) =\frac{3}{4\pi a^3(t)},\quad
m_h(t,r) = \frac{r^3}{a^3(t)},
\]	
and
\begin{equation} \label{homfield}
\dx U_h(t,x) = \frac{x}{a^3(t)},
\end{equation}
i.e., the macroscopic quantities related to this spatially homogeneous solution
$h_\epsilon$ do actually not depend on $\epsilon$.
It is well known that \eqref{eq:Definitiona} cannot be solved explicitly.
The following information on the behavior of $a$ will be useful.

\begin{lemma}\label{lem:PropertiesOfa}
Let $a \colon [0,T[ \to ]0,\infty[$ be the maximal solution of 
\eqref{eq:Definitiona}. Then $T=\frac{\pi}{2\sqrt{2}}{\mathring a\,}^{3/2}$,
$a$ is strictly decreasing on $[0,T[$
with $\lim_{t\to T} a(t)=0$, and for all $t\in ]0,T[$,
\begin{equation} \label{eq:DotaExplicitly}
\dot a(t) =-\sqrt{2}\sqrt{\frac{1}{a(t)}-\frac{1}{\mathring a}} < 0,
\end{equation}
\begin{equation} \label{eq:ArctanEquationFora}		
\frac{a(t)}{\mathring a}\sqrt{\frac{\mathring a}{a(t)}-1}
+ \arctan \sqrt{\frac{\mathring a}{a(t)}-1}
=\sqrt{2} \mathring a^{-3/2} t.
\end{equation}
\end{lemma}

\prf
As long as the solution exists, $\ddot a <0$ and hence
$\dot a (t) < \dot a(0)=0$ for $t>0$.
We multiply the differential equation in \eqref{eq:Definitiona} by $\dot a$
and integrate to find that
\begin{equation} \label{eq:IntegralOfaOde}
\frac{1}{2}\dot a^2(t)=\frac{1}{a(t)}-\frac{1}{\mathring a}
\end{equation}
on $[0,T[$ which yields \eqref{eq:DotaExplicitly}. 
Using the substitution 
\[
b=\sqrt{\frac{1}{a(t)}-\frac{1}{\mathring a}},
\]
this equation can be integrated once more to yield 
\eqref{eq:ArctanEquationFora}. Since $\lim_{t\to T} a(t)=0$,
the formula for $T$ is obtained by taking the corresponding limit in
\eqref{eq:ArctanEquationFora}, 
and the proof is complete.
\prfe

\begin{remark}
Let $r(t) = a(t)/{\mathring a},\ t\in [0,T[$. Then
\[
\rho(t,x) := \frac{3}{4\pi a^3(t)}\mathbf{1}_{B_{r(t)}(0)}(x),\ 
u(t,x):=\frac{\dot a(t)}{a(t)} x
\]
defines a solution of the pressure-less Euler-Poisson system
which coincides with our introductory collapse example 
if $\mathring a =1$.
\end{remark}

This solution can be viewed as follows. We start with the spatially
homogeneous solution with density $\rho_h$ and the given velocity field
and cut from it a spherically symmetric piece the boundary of which
is given by the curve $r=r(t)$. It should be noted that this curve
is precisely the trajectory of the particle which starts at radius 
$r=1$ with zero initial velocity. 

An analogous boundary curve for a corresponding cut in the
Vlasov case does not exist since there is at each point in space
a distribution of particles with different velocities.
We therefore proceed as follows. We choose a family
of cut-off functions 
$\phi_\epsilon \in C^\infty([0,\infty[)$, $\epsilon \in ]0,1]$,
such that
\[
0\leq \phi_\epsilon \leq 1,\ 
\phi_\epsilon(r)=1\ \mbox{for}\ r \leq 1,\
\phi_\epsilon(r)=0\ \mbox{for}\ r > 1 +\epsilon.
\]
The initial data
\begin{equation} \label{initialdata}
\mathring f_\epsilon (x,v)
:= h_\epsilon (0,x,v) \phi_\epsilon(|x|)
\end{equation}
launch a smooth, global, spherically symmetric solution 
$f_\epsilon$ of the Vlasov-Poisson system; notice that
initially this solution coincides with the homogeneous one on 
$B_1(0)\times \R^3$.
We aim to show that for $\epsilon$ small a homogeneous core at the
center persists arbitrarily closely up to the collapse time $T$
of the homogeneous solution from which fact Theorem~\ref{thm:Main} 
will follow. In order to define the boundary
of the homogeneous core, we place a point mass which is 
slightly larger than the total mass of the initial data  
at the origin and consider the trajectory of a particle which
moves radially inward in the corresponding potential and
starts at the cut-off radius 1  
with an initial radial velocity which in modulus is larger
than the initial radial velocities of the Vlasov particles.
To make this precise we define a strict upper bound
for the total mass of $f_\epsilon$ by
\begin{equation} \label{Mepsdef}
M_\epsilon := \iint \mathring f_\epsilon (x,v)\,dv\,dx + \epsilon
\end{equation}
and let $r_\epsilon\colon [0,T_\epsilon[ \to ]0,\infty[$ be the maximal 
solution of the initial value problem
\begin{equation} \label{eq:Definitionr_M}
\ddot r = -\frac{M_\epsilon}{r^2},\ 
r(0)=1,\ \dot r(0)=-\epsilon .
\end{equation}
We can now state our main result.
\begin{theorem} \label{thm:approx}
Let $f_\epsilon$ and $r_\epsilon$ be defined as above for 
$\epsilon\in ]0,1]$. Then the following holds. 
\begin{itemize}
\item[(a)]
$T_\epsilon < T$ for $\epsilon\in ]0,1]$ with
$T_\epsilon \to T$ for $\epsilon \to 0$.
\item[(b)]
$r_\epsilon(t) \leq  a(t)/\mathring a$ for $t\in [0,T_\epsilon[$
and $\epsilon\in ]0,1]$, and
$r_\epsilon(t) \to  a(t)/\mathring a$ for 
$\epsilon\to 0$, uniformly on any time interval $[0,T']\subset [0,T[$.
\item[(c)]
$f_\epsilon (t,x,v) = h_\epsilon(t,x,v)$ for $\epsilon\in ]0,1]$,
$t\in [0,T_\epsilon[$,
$|x| \leq r_\epsilon(t)$, and $v\in \R^3$.
\end{itemize}
\end{theorem}
The theorem will be proven in a number of steps in the next section.
We first indicate how it implies Theorem~\ref{thm:Main}. 

\smallskip

{\bf Proof of Theorem~\ref{thm:Main}.}
For all $\epsilon \in ]0,1]$ the following estimates hold at time
$t=0$. First of all, \eqref{initialdata} and the properties
of $H_\epsilon$ and $\phi_\epsilon$ imply that
\[
0\leq \rho(0,x) \leq \frac{3}{4\pi {\mathring a}^3} \mathbf{1}_{B_2(0)}(x),
\ x\in \R^3.
\]
This in turn implies that
\[
m(0,r) \leq  \frac{r^3}{{\mathring a}^3}\ \mbox{for}\ r\leq 2,\
m(0,r) \leq  \frac{8}{{\mathring a}^3}\ \mbox{for}\ r> 2.
\]
In particular,
\[
\sup_{r > 0}\frac{m(0,r)}{r} \leq  \frac{4}{{\mathring a}^3}.
\]
Moreover,
\[
E_\mathrm{kin}(0) 
\leq
\frac{1}{2}\frac{32\pi}{3}\int |v|^2 H_\epsilon ({\mathring a}^2|v|^2)\, dv 
=
\frac{16\pi}{3} 4\pi \int_0^1 s^4 H(s^2)\, ds \frac{1}{{\mathring a}^5}.
\]
Finally,
\[
-E_\mathrm{pot}(0) 
\leq  
\frac{1}{2} \left(\frac{3}{4\pi {\mathring a}^3}\right)^2
\int_{|x|\leq 2}\int_{|y|\leq 2}\frac{1}{|x-y|}dy\,dx.
\]
These estimates show that by choosing $\mathring a$ sufficiently large
the estimates at $t=0$ in Theorem~\ref{thm:Main} hold.

Consider now some $\epsilon \in ]0,1]$ and $0<t<T_\epsilon$.
Theorem~\ref{thm:approx}~(c) implies that
\[
\rho(t,x) = \frac{3}{4\pi a^3(t)},\ |x| \leq r_\epsilon(t).
\]
This in turn implies that
\[
m(t,r) = \frac{r^3}{a^3(t)},\ r\leq r_\epsilon(t)
\]
so that in particular
\[
\sup_{r>0}\frac{m(t,r)}{r} \geq \frac{r_\epsilon^2(t)}{a^3(t)}.
\]
Finally,
\begin{eqnarray*}
-E_\mathrm{pot}(t) 
&\geq&  
\frac{1}{2} \left(\frac{3}{4\pi a^3(t)}\right)^2
\int_{|x|\leq r_\epsilon(t)}\int_{|y|\leq r_\epsilon(t)}\frac{1}{|x-y|}dy\,dx\\
&=&
\frac{1}{2} \left(\frac{3}{4\pi}\right)^2
\int_{|x|\leq 1}\int_{|y|\leq 1}\frac{1}{|x-y|}dy\,dx 
\frac{r_\epsilon^5(t)}{a^6(t)}.
\end{eqnarray*}
Using parts (a) and (b) of Theorem~\ref{thm:approx} together with
the fact that $\lim_{t\to T} a(t) =0$, all these quantities can be made
large in the sense of Theorem~\ref{thm:Main} by making $\epsilon$
small and choosing $t$ close to $T$; when doing this $\mathring a$ and
hence the estimates at $t=0$ remain unchanged. The fact that the kinetic energy
behaves in the same way follows from conservation of energy,
and the proof is complete. \prfe

\section{Proof of Theorem~\ref{thm:approx}}
\setcounter{equation}{0}

We first observe that the parameter $\mathring a$ was used only to
make sure that the initial estimates in Theorem~\ref{thm:Main}
hold. Since it plays no role in the proof of Theorem~\ref{thm:approx}
we can for the rest of this paper
simplify our notation by choosing $\mathring a = 1$.

\subsection{Proof of parts (a) and (b) of Theorem~\ref{thm:approx}}

The initial data for the functions $a$ and $r_\epsilon$
imply that $r_\epsilon < a$ on some interval 
$]0,t^\ast[ \subset [0,T[\cap [0,T_\epsilon[$ where we choose $t^\ast >0$
maximal. The definition \eqref{Mepsdef} together with the properties
of $\phi_\epsilon$ imply that
\[
1 < M_\epsilon < (1+\epsilon)^3 + \epsilon.
\]
We use the lower bound to conclude from the differential equations
for $a$ and $r_\epsilon$ that on $]0,t^\ast[$ the estimate
$\ddot r_\epsilon < \ddot a$ holds, and since 
$\dot r_\epsilon(0)=-\epsilon < \dot a(0)$
also $\dot r_\epsilon < \dot a$. This implies that
$t^\ast = \min (T,T_\epsilon)$. Since the maximal existence time
$T$ respectively $T_\epsilon$ is determined by the fact that the
function $a$ respectively $r_\epsilon$ becomes zero there and
since the difference $a-r_\epsilon$ is positive and strictly increasing
as long as both functions exist we can conclude that
$T_\epsilon < T$ and $r_\epsilon < a$ on $]0,T_\epsilon[$.

As in the proof of Lemma~\ref{lem:PropertiesOfa}, we see that
$r_\epsilon$ is a strictly decreasing function with
$\dot r_\epsilon < -\epsilon$,
and 
\[
(\dot r_\epsilon(t))^2  - \epsilon^2 = 
2 M_\epsilon\left(\frac{1}{r_\epsilon(t)}-1\right).
\]
Hence
\[
\dot r_\epsilon(t) =-\sqrt{2 M_\epsilon}\sqrt{\frac{1}{r_\epsilon(t)}-C_\epsilon}
\quad\text{with}\quad C_\epsilon :=1-\frac{\epsilon^2}{2 M_\epsilon}.
\]
We note that $\lim_{\epsilon \to 0}M_\epsilon = 1$ and
$0 < C_\epsilon < 1$ with 
$\lim_{\epsilon \to 0}C_\epsilon =1$. Let us define 
a function $F\colon [0,1] \to \R$ by
\[
F(r) := r\sqrt{\frac{1}{r}-1}+\arctan \sqrt{\frac{1}{r}-1}\
\mbox{for}\ r\in ]0,1],\ F(0):=\frac{\pi}{2}.
\]
This function is continuous and differentiable on $]0,1[$ with
$F'(r) = -1/\sqrt{\frac{1}{r}-1} < 0$. 
Hence $F\colon [0,1] \to [0,\pi/2]$
is strictly decreasing and onto with a continuous inverse.
Using $F$, the above differential equation for $r_\epsilon$
can be integrated and yields the relation
\[
F(C_\epsilon r_\epsilon(t)) - F(C_\epsilon) 
= \sqrt{2 M_\epsilon C_\epsilon^3}\, t,\ t\in [0,T_\epsilon[.
\]
Since $\lim_{t\to T_\epsilon} r_\epsilon(t) = 0$ 
and $\lim_{r\to 0} F(r)=\pi/2$, it follows that
\[
T_\epsilon = \frac{1}{\sqrt{2 M_\epsilon C_\epsilon^3}}
\left(\frac{\pi}{2} - F(C_\epsilon)\right) \to \frac{\pi}{2\sqrt{2}} = T
\]
for $\epsilon \to 0$ as claimed.
Now fix some $t\in [0,T[$. Then for $\epsilon$ sufficiently small,
$t\in [0,T_\epsilon[$, and
\[
r_\epsilon(t) = \frac{1}{C_\epsilon} 
F^{-1}\left(F(C_\epsilon) + \sqrt{2 M_\epsilon C_\epsilon^3}\, t\right)
\to F^{-1} \left(\sqrt{2}\, t\right)  = a(t)
\]
as $\epsilon \to 0$; here we used the limit behavior of
$M_\epsilon$ and $C_\epsilon$ and the identity \eqref{eq:ArctanEquationFora}.
This proves the desired limit for $r_\epsilon(t)$, and since
$0\leq a(s) - r_\epsilon(s) \leq a(t) - r_\epsilon(t)$ on $[0,t]$
the limit is uniform on compact subintervals of $[0,T[$.
Parts (a) and (b) of Theorem~\ref{thm:approx} are proven. \prfe

\subsection{The behavior of characteristics}

To prove part (c) of Theorem~\ref{thm:approx},
we use the fact that a smooth function $f$ solves the Vlasov
equation \eqref{vlasov} if and only if it is constant along its characteristics,
i.e., along the solutions of the characteristic system
\begin{equation} \label{CS}
\dot x = v,\ \dot v = -\partial_x U(s,x).
\end{equation}
In particular, if $[0,\infty[ \ni s \mapsto (X(s,t,x,v),V(s,t,x,v))$
denotes the solution of the characteristic system with initial data
$(X(t,t,x,v),V(t,t,x,v))=(x,v)$ with $t\geq 0$ and $x,v \in \R^3$
prescribed, then a solution of the Vlasov-Poisson system $f$ is
related to its initial data $\mathring f$ by the relation
\[
f(t,x,v) = \mathring f(X(0,t,x,v),V(0,t,x,v)),\ t\in \R,\
x,v \in \R^3.
\]
To show that the solution $f_\epsilon$ launched by $\mathring f_\epsilon$ 
has a homogeneous core bounded by the curve $r=r_\epsilon$, 
we need to control characteristics which cross this curve.
This analysis will be facilitated by the spherical symmetry
of the solutions in question. In order to exploit this symmetry,
we define for $x,v\in \R^3$ corresponding spherical variables by
\begin{equation} \label{rwL}
r=|x|,\ w=\frac{x\cdot v}{r},\ L= |x\times v|^2.
\end{equation}
If $(x(s),v(s))$ solves \eqref{CS}, then in spherical variables,
\[
\dot r = w,\ \dot w = \frac{L}{r^3} - \frac{m(s,r)}{r^2},\ \dot L=0;
\]
spherical symmetry of the gravitational field implies that angular
momentum and also its square $L$ is constant along particle trajectories.

We first establish some control on characteristics 
under the assumption that
the gravitational field is generated by a mass distribution
of total mass bounded by $M_\epsilon$. 

\begin{lemma}\label{lem:ControlOfOutsideCharacteristics}
Let $\partial_x U = \partial_x U(t,x)= m(t,r)x/r^3$ be a spherically symmetric
and continuously differentiable gravitational field defined on 
$[0,T[ \times \R^3$
with the property that $0\leq m(t,r) < M_\epsilon$ for $t\in[0,T[$, $r\geq 0$,
and some $\epsilon\in ]0,1]$.	
Let $[0,T[ \ni s \mapsto (x(s),v(s))$ be a solution of the 
corresponding characteristic system 
\eqref{CS} with spherical representation $(r(s),w(s),L)$ as described in 
\eqref{rwL}.
\begin{itemize}
\item[(a)]
If $r(0) \geq 1$ and $|v(0)| < \epsilon$,
then $r(s)>r_\epsilon(s)$ for all $s\in ]0,T_\epsilon[$.
\item[(b)]
If $r(0) \leq 1$ and $|r(t)| \geq r_\epsilon(t)$ for some $t\in ]0,T_\epsilon[$,
then $r(s)>r_\epsilon(s)$ for all $s\in ]t,T_\epsilon[$.
\end{itemize}
\end{lemma}

\prf
We first consider part (a). 
Since by assumption $r(0)\geq 1=r_\epsilon(0)$ and 
$\dot r(0) > -\epsilon = \dot r_\epsilon(0)$, there exists 
$t\in ]0,T_\epsilon]$ such that $r(s) > r_\epsilon(s)$ for $s\in ]0,t[$,
and we choose $t$ maximal.
On the interval $]0,t[$,
\begin{equation}\label{ddotr}
\ddot r = \frac{L}{r^3}- \frac{m(s,r)}{r^2} > -\frac{M_\epsilon}{r^2}
> -\frac{M_\epsilon}{r_\epsilon^2} = \ddot r_\epsilon.
\end{equation}
If $t<T_\epsilon$ we use the assumptions at $s=0$ and 
integrate \eqref{ddotr} twice
to find that $r(t) > r_\epsilon(t)$ which contradicts the maximality of
$t$. Hence $t=T_\epsilon$, and part~(a) is proven.

As to part~(b) we first show that there exists a time $t'\in ]0,t]$
such that 
\begin{equation}\label{tprime}
r(t') \geq r_\epsilon(t')\ \mbox{and}\ w(t')\geq \dot r_\epsilon(t').
\end{equation}
This can be seen as follows. If  $w(t)\geq \dot r_\epsilon(t)$
we choose $t'=t$. If  $w(t)< \dot r_\epsilon(t)$  there exists
$t^\ast \in [0,t[$ such that $r(s) > r_\epsilon(s)$ for $s\in ]t^\ast,t[$,
and we choose $t^\ast$ minimal. 
Assuming that $w(s)<\dot r_\epsilon(s)$ on  $]t^\ast,t[$
it would follow that $r(t^\ast) > r_\epsilon(t^\ast)$.
If $t^\ast >0$, this 
contradicts the minimality of $t^\ast$, and if $t^\ast =0$
it contradicts the assumption $r(0) \leq 1 = r_\epsilon(0)$ in part~(b).
Hence there must exist
a time $t'$ such that \eqref{tprime} holds.
For any time $s\in [t',T_\epsilon[$ such that $r(s) \geq r_\epsilon(s)$
it follows that $\ddot r(s) > \ddot r_\epsilon(s)$, cf.\ \eqref{ddotr}.
In particular, this holds for $s=t'$. Hence there exists 
$t^\ast \in ]t',T_\epsilon]$ such that $r(s) > r_\epsilon(s)$
for $s\in ]t',t^\ast[$, and we choose $t^\ast$ maximal.
Assuming $t^\ast <T_\epsilon$ we integrate  
the inequality $\ddot r(s) > \ddot r_\epsilon(s)$ twice starting
at $t'$ and using the properties \eqref{tprime} to conclude that
$r(t^\ast) > r_\epsilon(t^\ast)$ in contradiction to the maximality of $t^\ast$.
Hence $t^\ast = T_\epsilon$, and the proof of part (b) is complete.
\prfe

Next we consider the characteristics of the homogeneous solution $h_\epsilon$.
\begin{lemma}\label{lem:ControlOfHomogenousCharacteristics}
Let $\epsilon \in ]0,1]$ and
let $[0,T[ \ni s \mapsto (x(s),v(s))$ be a characteristic curve
of the homogeneous solution $h_\epsilon$, i.e., in 
\eqref{CS} the field $\partial_x U$ is given by \eqref{homfield},
with spherical representation $(r(s),w(s),L)$ as described in 
\eqref{rwL}. If 
\[
t\in [0,T_\epsilon[\quad \mbox{with}\quad 
h_\epsilon (t,x(t),v(t)) > 0\quad\text{and}\quad r(t) \leq r_\epsilon(t),
\]
then 
$r(s) < r_\epsilon(s)$ for all $s\in[0,t[$.
\end{lemma}

\prf
Since we want to use the spherical representation
of the given characteristic, we first assume that
$x(s) \neq 0$ for $s \in ]0,t[$.
In order to exploit the fact that the
characteristics of the homogeneous solution $h_\epsilon$
remain close to the trajectories of the corresponding dust particles
it is convenient 
to rewrite the characteristic equations in coordinates which
are co-moving with the particles of the corresponding homogeneous
dust solution, i.e.,
\begin{eqnarray*}
\tilde r(s) 
&:=& 
\frac{r(s)}{a(s)},\\
\tilde w(s) 
&:=& 
a^2(s) \dot {\tilde r}(s) = a(s) w(s) - \dot a(s) r(s).
\end{eqnarray*} 
The separating curve is transformed accordingly, i.e.,
\begin{eqnarray*}
\tilde r_\epsilon(s)
&:=& \frac{r_\epsilon(s)}{a(s)},\\
\tilde w_\epsilon(s)
&:=&
a^2(s)\dot{\tilde r}_\epsilon(s) 
= a(s) \dot r_\epsilon(s) - \dot a(s) r_\epsilon(s).
\end{eqnarray*}
Using the already established part (b) of Theorem~\ref{thm:approx}
and recalling that we took $\mathring a = 1$ we find that on
the interval $]0,T_\epsilon[$,
\[ 
\dot{\tilde w}_\epsilon 
=
a \, \ddot r_\epsilon - \ddot a\, r_\epsilon
= \frac{r_\epsilon}{a^2} - a\frac{M_\epsilon}{r_\epsilon^2}
\leq
\frac{1}{a} - \frac{M_\epsilon}{a}=\frac{1}{a}(1-M_\epsilon)<0.
\]
Thus $\tilde w_\epsilon(s) < \tilde w_\epsilon(0)=-\epsilon$, and hence
\begin{equation}\label{eq:ControlOfDotTilder_M}
\dot{\tilde r}_\epsilon (s) <-\frac{\epsilon}{a^2(s)},\ s\in ]0,T_\epsilon[. 
\end{equation}
To compare this to the given, mass-carrying characteristic of the 
homogeneous solution,
we observe that by the definition \eqref{eq:DefinitionHepsilon} of 
$H_\epsilon$, 
\begin{eqnarray*}
\epsilon^2 
&>& 
|a(s)v(s)-\dot a(s)x(s)|^2 \\
&=& 
a^2(s)|v(s)|^2-2a(s)\dot a(s)(x\cdot v)(s)+\dot a^2(s)|x(s)|^2\\
&=& 
a^2(s)\frac{L}{r^2(s)} +\left(a(s)w(s)-\dot a(s)r(s)\right)^2\\
&=& 
\frac{L}{\tilde r^2(s)}+\tilde w^2(s).
\end{eqnarray*}
Thus for $s\in ]0,t[$ it follows that
$\tilde w(s) >-\epsilon$ which by definition of $\tilde w$ and
\eqref{eq:ControlOfDotTilder_M} implies that
\[
\dot{\tilde r}(s) >-\frac{\epsilon}{a^2(s)}>\dot{\tilde r}_\epsilon (s).
\]
Since $\tilde r(t)\le \tilde r_\epsilon(t)$, it follows that
$\tilde r(s) < \tilde r_\epsilon(s)$ and
hence $r(s)<r_\epsilon(s)$ on $[0,t[$ as desired.

So far we assumed that $x(s)\neq 0$ on the interval
$]0,t[$. For times $s\in [0,t[$ where $x(s)= 0$ the assertion 
of the lemma holds since $r_\epsilon(s)>0$. If
the function $x(s)$ has zeros but is not identically zero,
we let $[s_1,s_2] \subset [0,t]$ be an interval with non-empty
interior and such that $x(s)\neq 0$ on $]s_1,s_2[$; we choose
the interval maximal with this property. 
Then $x(s_2)=0$ or $s_2=t$, and in either case
$r(s_2) \leq r_\epsilon(s_2)$ so that the above argument
on the interval $[0,t]$ now applies to $[s_1,s_2]$
and implies that $r(s)<r_\epsilon(s)$ on 
$[s_1,s_2[$.
Since the interval $[0,t[$ is the union of such subintervals
and a set of points where $x(s)=0$,  
the proof is complete. 
\prfe

\subsection{Proof of part (c) of Theorem~\ref{thm:approx}}

Using the above information on characteristics,
we can prove the remaining assertion of Theorem~\ref{thm:approx}.
Since in these arguments the parameter $\epsilon$
remains fixed, we write $f=f_\epsilon$ for the solution
of the Vlasov-Poisson system launched by the initial
data $\mathring f=\mathring f_\epsilon$ specified in \eqref{initialdata}
and $h$ for the homogeneous solution. 
We recall that
$\iint \mathring f < M_\epsilon$,
\[
\mathring f(x,v) =h(0,x,v),\quad (x,v)\in B_1(0)\times \R^3,
\]
and
\[
\mathring f(x,v)= 0\quad\text{if}\quad 
|v| \geq \epsilon;
\]
the latter follows from the definition \eqref{hdef}
of the homogeneous solution $h$ and \eqref{initialdata}.
We define
\[
I:= 
\left\{(t,x,v)\in [0,T_\epsilon[\times\R^3\times\R^3\mid 
|x| < r_\epsilon(t)\right\},
\]
and we have to show that $f|_I = h|_I$. To prove this we first
establish the following assertion for these functions on the boundary
of $I$:
\begin{eqnarray} \label{bcI}
\forall t\in ]0,T_\epsilon[,\ x,v \in \R^3\ \mbox{with}\
|x|=r_\epsilon(t)&& \mbox{and}\ w=\frac{x\cdot v}{r}\leq \dot r_\epsilon(t):
\nonumber \\
&&
f(t,x,v)=0=h(t,x,v).
\end{eqnarray}
To prove the assertion for $f$ we consider the characteristic curve
$(x(s),v(s))=(X,V)(s,t,x,v)$ of $f$ which at time $s=t$ passes through
a boundary point as specified in \eqref{bcI} with $w<\dot r_\epsilon(t)$.
Then there exists some $\delta>0$ such that 
$|x(s)| < r_\epsilon (s)$ for $t<s<t+\delta$ and
$|x(s)| > r_\epsilon (s)$ for $t-\delta<s<t$. 
Hence Lemma~\ref{lem:ControlOfOutsideCharacteristics}~(b) implies that
$|x(0)| >1$, and Lemma~\ref{lem:ControlOfOutsideCharacteristics}~(a) 
implies that $|v(0)|\geq \epsilon$
and hence $f(t,x,v)=\mathring f(x(0),v(0))=0$; by continuity the assertion
also holds if $w = \dot r_\epsilon(t)$.
In order to prove  \eqref{bcI} for $h$ we argue in the same way
using a characteristic curve of $h$, and 
Lemma~\ref{lem:ControlOfHomogenousCharacteristics} implies that
$h(t,x,v)=0$.

The idea now is that there can be at most one solution
of the Vlasov-Poisson system on $I$ which has given data at $t=0$
and satisfies the boundary condition \eqref{bcI}.
If we take the difference of the Vlasov equations for $f$ and $h$,
we find that
\begin{equation} \label{vl_f-h}
\partial_t (f-h) + v\cdot\partial_x(f-h) - 
\partial_x U_f \cdot \partial_v (f-h) = \partial_x(U_f - U_h) \cdot \partial_v h
\end{equation}
which holds for all $t\in [0,T_\epsilon[$ and $x,v \in \R^3$;
$U_f$ and $U_h$ denote the potentials induced by $f$ respectively $h$.
We consider a characteristic curve 
$(x(s),v(s))=(X,V)(s,t,x,v)$ of $f$ with $|x| < r_\epsilon(t)$
and define
\[
s^\ast := \sup\left\{ s\in [0,t] \mid 
|x(\tau)| < r_\epsilon (\tau),\ s\leq \tau \leq t\right\}
\]
so that $|x(s)| < r_\epsilon (s)$ on $]s^\ast,t]$. Integrating
\eqref{vl_f-h} along the characteristic curve implies that
\begin{eqnarray}\label{vlint_f-h}
(f-h)(t,x,v)
&=&
(f-h)(s^\ast,x(s^\ast),v(s^\ast)) \nonumber \\
&&
{}+
\int_t^{s^\ast}
\left(\partial_x(U_f-U_g)\cdot \partial_v h\right)(s,x(s),v(s))\, ds
\nonumber\\
&=&
\int_t^{s^\ast}
\left(\partial_x(U_f-U_g)\cdot \partial_v h\right)(s,x(s),v(s))\, ds;
\end{eqnarray}
notice that either $s^\ast=0$ in which case the first term on the 
right hand side vanishes
because both functions have the same initial data
for $|x| \leq 1$, or  $s^\ast>0$ in which case $|x(s^\ast)|= r_\epsilon(s^\ast)$
and $w(s^\ast) \leq \dot r_\epsilon(s^\ast)$, and the first term on the 
right hand side vanishes
due to \eqref{bcI}.
For $t\in [0,T_\epsilon[$ we define 
\[
D(t):= \sup\{|f-h|(t,x,v) \mid |x|\leq r_\epsilon (t),\ v\in \R^3\}.
\]
Spherical symmetry and standard estimates imply that
\[
\sup\{|\partial_x U_f-\partial_x U_h|(t,x) \mid |x|\leq r_\epsilon (t)\}
\leq C D(t);
\]
notice that the velocity supports of both $f$ and $h$ are
bounded, uniformly on $[0,T_\epsilon[$, and hence
$|\rho_f(t,x)-\rho_h(t,x)| \leq C D(t)$ for $|x|\leq r_\epsilon (t)$. 
Hence \eqref{vlint_f-h} implies that
\[
D(t) \leq C \int_0^t D(s)\, ds
\]
so that $D(t)=0$ for $t\in [0,T_\epsilon[$,
and the proof of Theorem~\ref{thm:approx}
is complete.


\begin{thebibliography}{0}
\bibitem{BT}
J.~Binney, S.~Tremaine,
\textit{Galactic Dynamics}, Princeton University Press 1987.

\bibitem{DS} 
C.~Dietz, V.~Sandor, 
\textit{The hydrodynamical limit of the Vlasov-Poisson system}, 
Transport Theory Statist.\ Phys.\ 
\textbf{28} (1999), 499--520.

\bibitem{LP}
P.-L.~Lions, B.~Perthame,
\textit{Propagation of moments and regularity for the 3-dimensional
Vlasov-Poisson system},
Invent.\ Math.\ \textbf{105} (1991), 415--430.

\bibitem{OS} 
J.~R.~Oppenheimer, H.~Snyder, 
\textit{On continued gravitational contraction},
Physical Review \textbf{56} (1939), 455--459.

\bibitem{Pf}
K.~Pfaffelmoser,
\textit{Global classical solutions of the Vlasov-Poisson system in three
dimensions for general initial data},
J.\ Differential Equations\
\textbf{95} (1992), 281--303.

\bibitem{Rein07}
G.~Rein,
\textit{Collisionless Kinetic Equations from Astrophysics---The Vlasov-Pois\-son System},
Handbook of Differential Equations, Evolutionary Equations. \textbf{3} (2007),
Eds.\ C.~M.~Dafermos and E.~Feireisl, Elsevier.

\bibitem{RR}
G.~Rein, A.~D.~Rendall,
{Global existence of classical solutions to the Vlasov-Poisson system
in a three-dimensional, cosmological setting},
Arch.\ Rational Mech.\ Anal.\
\textbf{126} (1994), 183--201.

\bibitem{diss}
L.~Taegert,
{\em Oppenheimer-Snyder type collapse with Vlasov matter}, 
Doct\-oral thesis, Bayreuth, in preparation.
\end{thebibliography}
\end{document}